# AdaptiveDeltaXAI: Adaptive Online Explanation for Continuous Datacenter Telemetry Monitoring

Richard T. Rajkumar, Ridwan Amure, Muzzakir Mohammed, Mariofanna Milanova, Abdelrahman Elfikky
*School of Computer and Information Sciences, University of Arkansas at Little Rock, AR 72204 USA.*

## Abstract

Continuous monitoring repeatedly raises a prediction-change question: *why did the anomaly score change since the previous observation?* Delta-XAI [1] answers it with Full-SWING, a fresh prediction-change attribution, but recomputing it at every step is costly for continuous datacenter telemetry. We build on Delta-XAI and do not propose a new attribution method. AdaptiveDeltaXAI places an adaptive refresh controller around fresh Full-SWING attribution and labels reused maps as cached event context. The controller combines an EMA score-residual trigger, input-drift and maximum-age safeguards, and display-only temporal discounting; predefined feature groups are produced separately. On a controlled 48-hour synthetic stream, three-seed full-stream refresh rates are 8.09% ± 0.02% for temperature and 4.63% ± 0.25% for power. A repeated T4 batched probe reduces measured explanation-stage latency from 62.59 ± 6.49 to 9.75 ± 0.24 ms per decision, a 6.42-fold ratio of means; routine detector scoring and grouping are excluded. At retrospectively matched public refresh counts, adaptive event-refresh coverage is 100% across three named SMD units and 75.0% on PSM, compared with 82.8%/52.8% for periodic and 86.1%/40.3% for random controls. A direct prediction-difference perturbation test on 50 common temperature targets does not support Full-SWING superiority. These results support workload-specific scheduling savings while separating fresh-map completeness, event coverage, and attribution faithfulness. Code, tests and results tables: github.com/richardtrajkumar/AdaptiveDeltaXAI (release v1.9).

## 1. INTRODUCTION

When an on-call engineer receives an alert at 3AM because an anomaly score crossed its threshold, the next question is why the score changed and which sensor channels contributed. Continuous datacenter monitoring poses this question repeatedly: a new score arrives at each sampling interval, while a fresh gradient-based explanation has a nonzero computational cost. We study how to schedule these fresh explanations and make the meaning of a reused display explicit.

Modern datacenter monitoring systems use machine-learning models to transform continuous telemetry into anomaly scores or failure-risk probabilities. These models may include statistical detectors, autoencoders, recurrent neural networks, convolutional sequence models, or transformer-based time-series models. They consume multivariate windows of temperature, power, fan, voltage, utilization, and environmental measurements and produce a scalar score, typically compared with an alert threshold. Although the score identifies potentially abnormal behavior, it does not identify the underlying source. An unexplained alert can delay triage, direct an engineer toward healthy equipment, or obscure an emerging cooling, power, or hardware failure.

This limitation makes explanation of prediction changes essential. A current anomaly probability of 0.80 can describe a rapidly worsening event, a stable condition already under investigation, or a recovering system whose prior score was higher. These states have different operational meanings even though the current prediction is identical. An operator therefore needs an explanation of the transition from the prior prediction to the current prediction: which sensor channels changed, how they contributed to the model's changing assessment, and whether the change creates a new reason to act.

Delta-XAI [1] addresses this distinction by explaining differences between time-series predictions rather than explaining each prediction independently. Its Shifted Window Integrated Gradients (SWING) method attributes the change between adjacent prediction windows to the input features responsible for that change. Our focus is the operational cost of repeatedly applying prediction-change attribution to fixed-interval, multivariate datacenter telemetry.

This creates a systems-level XAI problem. Applying a fresh multipath attribution at every timestep incurs repeated gradient computation even when the controller does not require new event context. The measured cost depends jointly on the detector architecture, window length, feature count, quadrature, batching, and hardware. Normal telemetry also contains periodic behavior; the experiments therefore evaluate a score-residual trigger and safeguards rather than claim a universal periodicity-removal property.

We introduce AdaptiveDeltaXAI, an adaptive online explanation framework for continuous datacenter telemetry monitoring. AdaptiveDeltaXAI retains Full-SWING as its fresh prediction-change attribution mechanism and makes a separate operational decision about when to compute a fresh map. On cache hits, it returns explicitly labeled cached event context rather than a claim about the current prediction change. The framework combines A1 adaptive triggering, A2 EMA-based residual processing, A3 input-drift and maximum-age invalidation, A4 display-only temporal discounting, and A5 operational feature-group presentation. Our research question is whether fresh prediction-change explanations can be scheduled efficiently while preserving the completeness of fresh maps and retaining operationally useful event context between refreshes. We evaluate this question on a 48-hour controlled synthetic stream, three named Server Machine Dataset (SMD) [2] units, and Pooled Server metrics dataset (PSM) [3]; the latter public study is a documented subset, not an all-28-SMD claim.

## 1.1 Related Work

Classical post hoc XAI methods, including LIME [4], SHAP [5], and Integrated Gradients [6], explain the contribution of input features to a single model prediction. Time-series XAI extends these ideas through instance-wise temporal importance [7], temporal-dependency analysis [8], dynamic-mask explanations [9], and learned perturbations [10], as well as surrogate consistency, contrastive perturbation, gradients, or Shapley-value estimation, including ContraLSP [11], TimeX [12], TimeX++ [13], TIMING [14], and TimeSHAP [15]. ShaTS [16] incorporates a priori temporal and sensor/process grouping into Shapley-based explanations for industrial time series. These methods address temporal feature relevance and multivariate sensor interpretation but generally do not ask whether a full explanation should be recomputed at every point in a continuous stream.

Delta-XAI [1] is most closely related because it explains prediction differences rather than isolated predictions; DeltaSHAP [17] considers a related Shapley-value formulation. AdaptiveDeltaXAI uses a Full-SWING implementation that follows the published near/far multipath construction and passes internal structural and linear-completeness tests. The present work addresses the online refresh-scheduling problem, rather than claiming a universal attribution mechanism.

The contributions of this paper are as follows:

1. We formulate continuous datacenter telemetry explainability as a prediction-change attribution problem under a desired per-timestep sampling budget (Sections 2 and 3), while explicitly distinguishing a fresh explanation from cached event context.
2. We introduce AdaptiveDeltaXAI (Section 4), an online controller around fresh Full-SWING prediction-change attribution [1]. Its five components, A1-A5, govern fresh decisions, cache staleness, cached-context display, and operational feature-group presentation.
3. We evaluate a controlled 48-hour synthetic telemetry design with temperature and power channels and burst, overlapping, and gradual-drift fault scenarios (Sections 3 and 6.1). We further evaluate selected properties on three named SMD units [2] and PSM [3], with reconstruction labels aligned to the final observation of each scored window.
4. We compare five fresh attribution methods on 50 common held-out temperature targets using direct prediction-difference perturbations, and report a separate current-score masking proxy and measured T4 latency distributions (Sections 5.2 and 6.3). We evaluate scheduling through refresh rate, fresh-only completeness, measured explanation-stage latency, and public matched-budget controls. These experiments do not establish Full-SWING as the most faithful method under the tested perturbations.
5. We analyze the adaptive components through three-seed ablations, repeated batching-versus-caching timing experiments, and controlled 8/16/32/64-feature scaling (Sections 6.4 and 6.5). We also evaluate the synthetic controller with GRU [18] and standard Transformer [19] detectors under the same chronological split and experimental protocol (Section 6.7).

## 2. PROBLEM SETUP

### 2.1 Continuous Datacenter Monitoring System

We consider $D$ telemetry channels sampled every $\Delta t$ seconds. At observation endpoint $t$, the system receives a sensor vector with one value per channel:

$$x_t = \left(x_{t,1}, \dots, x_{t,D}\right)^T \in R^D \quad (1)$$

Each channel represents a measurement such as inlet temperature, CPU temperature, memory temperature, exhaust temperature, fan behavior, power-supply input, or power-supply output. These channels are physically related rather than independent. For example, a workload increase may raise power consumption, increased power consumption may raise CPU temperature, and thermal changes may then propagate to exhaust sensors. Cooling-system behavior can affect several temperature channels simultaneously.

At endpoint $t$, the detector receives the most recent $W$ observations, $X_t = [x_{t-W+1}, \dots, x_t]$. In the synthetic classifier protocol, this window is paired with the next-observation label $y_{t+1}$. Thus, the score $s_t$ is a one-step-ahead anomaly score: it is computed after $x_t$ is available and is evaluated against $y_{t+1}$. The end-exclusive notation used in the train/test loader is equivalent: $X^{(i)} = x[i - W{:}\, i]$ is paired with $y_i$, where $t = i - 1$.

$$X_t = [x_{t-W+1}, \dots, x_t] \in R^{W \times D} \quad (2)$$

A differentiable anomaly-scoring model $f$ maps $X_t$ to a scalar score $s_t$. For the synthetic classifier, $s_t = p_t \in [0,1]$ is the probability used to predict the next label $y_{t+1}$; for the public reconstruction detector, $s_t$ is a log1p reconstruction-MSE score aligned to the final observation of its end-exclusive reconstruction window and need not lie in [0, 1].

$$s_t = f(X_t) \in R \quad (3)$$

When a detector score meets or exceeds its alert threshold $\tau$, the system may notify an operator. AdaptiveDeltaXAI operates above a differentiable detector and does not replace its training or alerting logic. Synthetic experiments use GRU [18] and Transformer [19] classifiers. Public experiments use a two-layer GRU encoder and a feedforward head that reconstructs the final observation of each window; its scalar score is log1p of the mean squared reconstruction error. This design is inspired by recurrent reconstruction-based detection [20], but is not an implementation of that paper's LSTM sequence decoder.

The synthetic stream contains 17,280 observations and is divided chronologically into 12,096 training, 1,728 validation, and 3,456 test observations. Training observations alone determine normalization. Split-local detector windows pair $x[i - 20{:}\, i]$ with $y_i$, yielding 3,436 test windows and 331 positive targets. Detector event metrics consequently cover 12 contiguous positive intervals. The continuous-controller refresh analysis covers all 3,456 held-out target indices and 13 positive intervals, including one interval outside the split-local detector window set. Operational full-stream rates use 17,235 decisions; their final decision has no valid next-observation label and is excluded from label-dependent evaluation.

Figure 1 shows a held-out excerpt of the v1.9 temperature detector score and fresh-refresh decisions from the continuous full-stream controller. This view does not show anomaly labels. Cached event context is not a new attribution of the current transition.

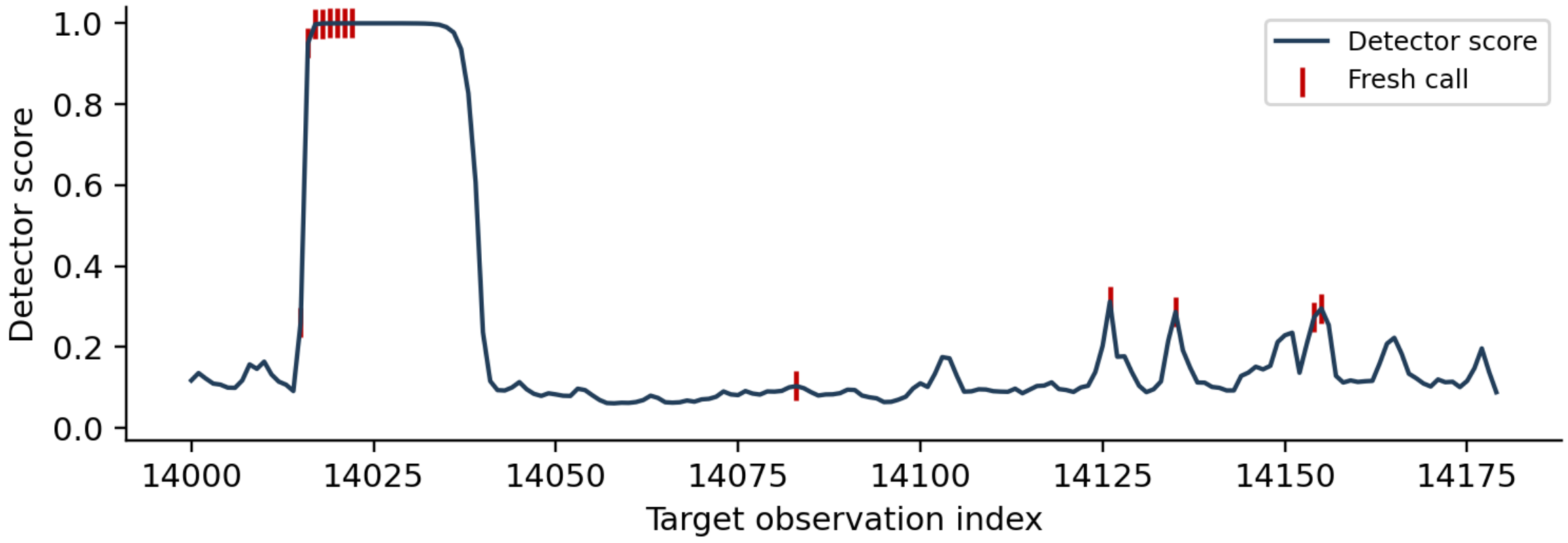

Figure 1: Held-out temperature detector score and fresh-refresh decisions for target indices 14000–14179 in the seed-42 continuous monitoring pass. Markers indicate fresh explanations; anomaly labels are not shown.

## 2.2 Dynamic XAI Problem Formulation

A conventional XAI method explains the model output for a single input. In continuous telemetry monitoring, the operational question is often why the detector score changed relative to the preceding window.

Let $X_t = [x_{t-W+1}, \ldots, x_t]$ denote the telemetry window available at endpoint $t$, where $W$ is the number of observations in the sliding window and $D$ is the number of sensor channels. In the synthetic classifier protocol, $X_t$ is paired with the next-observation target $y_{t+1}$. Let $X_{t-1}$ denote the preceding window. A differentiable anomaly-scoring model maps each window to a scalar score as in Equation (4).

$$s_t = f(X_t) \quad (4)$$

Let the score for the preceding window be $s_{t-1} = f(X_{t-1})$. The observed score change is then defined as

$$\Delta s_t = s_t - s_{t-1} = f(X_t) - f(X_{t-1}) \quad (5)$$

The prediction-change attribution assigns signed contributions to sensor channels. Let $H_t = \{t - W, \ldots, t\}$ be the absolute-time union of the two adjacent windows. Define the channel aggregate from the raw fresh map $E$ by

$$e_t = \left(e_{t,1}, \ldots, e_{t,D}\right)^T, \quad e_{t,d} = \sum_{l \in H_t} \quad E_{t,l,d} \quad (6)$$

The aligned raw fresh map $E$ has $W + 1$ temporal positions and $D$ channels. Equation (7) expresses numerical completeness: its total approximately equals the score difference. The corresponding synthetic targets are $y_t$ for the previous window and $y_{t+1}$ for the current window. Completeness is assessed only for fresh maps; it is not asserted for cached or discounted displays.

$$\sum_{d=1}^{D} \quad e_{t,d} = \sum_{l \in H_t} \quad \sum_{d=1}^{D} \quad E_{t,l,d} \approx \Delta s_t \quad (7)$$

For comparison, lag-wise subtraction of adjacent windows gives Equation (8). This subtraction compares relative window positions; it is not the absolute-time alignment used for Full-SWING accumulation.

$$\Delta X_t = X_t - X_{t-1} \quad (8)$$

For a fresh call, Full-SWING [1] uses four path terms: $A_t^F = IG(X_t, B^F; P^+)$, $A_t^N = IG(X_t, B^N)$, $A_{t-1}^F = IG(X_{t-1}, B^F)$, and $A_{t-1}^N = IG(X_{t-1}, B^N; P^-)$. $IG(A, B; P)$ integrates from baseline B to A along the indicated piecewise-linear path. Each term is a $W \times D$ map in relative window coordinates. $J_t$ and $J_{t-1}$ embed current and preceding maps into their respective positions on the common $(W + 1) \times D$ absolute-time union. The implemented composition is

$$E_t = \frac{1}{2} J_t(A_t^F + A_t^N) - \frac{1}{2} J_{t-1}(A_{t-1}^F + A_{t-1}^N) \quad (9)$$

The alignment operators in Equation (9) are necessary: subtracting the two $W \times D$ arrays directly would mix different absolute observation times. Appendix C specifies all operands and the quadrature. The frozen configuration uses four path terms and 64 trapezoidal segments, hence 65 gradient nodes per term. Internal structural and linear-completeness checks are recorded in the frozen precheck; numerical parity with a pinned external implementation is not established.

Let $L_t$ denote explanation-stage latency at decision $t$. The sampling interval supplies a desired budget, not a worst-case execution guarantee. The controller seeks lower average cost while returning a fresh current-change map on refresh steps and labeled cached event context on other steps. Equation (10) compares latency and sampling interval after expressing both in the same time unit.

$$L_t \leq \Delta t \quad (desired\ budget) \quad (10)$$

The Dynamic XAI problem is therefore to schedule fresh explanations under a desired sampling budget, while reporting observed fresh-call and amortized latency. The controller does not provide a worst-case execution-time guarantee.

## 2.3 Existing Approaches and Their Limitations

A direct solution is to apply an XAI method independently at every timestep. LIME [4], KernelSHAP [5], and Integrated Gradients [6] ordinarily explain a single prediction; Full-SWING [1] directly targets a prediction change.

$$Work_{fresh} \propto N_{path} N_{node} C_{node} \quad (11)$$

We report fresh-call latency and amortized explanation-stage latency separately, and isolate batching from scheduling in a repeated 2 × 2 experiment. In the saved implementation, detector scores are computed before the per-step timer; operational grouping is produced after controller execution. The measured controller latency therefore excludes routine detector inference and group presentation. It includes refresh decisions, state updates, cache handling, and any fresh Full-SWING call, with one initialization-time preceding-score evaluation. It is not a complete telemetry-to-interface latency. In Equation (11), $C_{node}$ denotes detector forward/backward work per gradient node; $N_{path} = 4$ and $N_{node} = 65$. Batching changes execution latency rather than this node count.

Unconditional recomputation pays for a fresh attribution at every score transition, including stable periods. A raw-change trigger can also react to periodic telemetry that is normal operating behavior. We therefore evaluate an adaptive score statistic and explicit staleness safeguards, without assuming that every skipped explanation is redundant or that every refresh corresponds to a fault.

Simple caching alone does not fully address these limitations. A previously computed explanation may become stale when a gradual fault develops, when a new sensor begins contributing to an existing event, or when the propagation pattern of a multichannel anomaly changes. An online explanation method must therefore balance two competing requirements: it should avoid unnecessary recomputation when the explanatory state is stable, while still producing a fresh explanation when meaningful changes occur.

These limitations motivate the adaptive recomputation strategy introduced in Section 2.4.

## 2.4 AdaptiveDeltaXAI Approach

Delta-XAI provides the attribution mechanism used to explain changes between consecutive model predictions. However, as discussed in Sections 2.2 and 2.3, continuous telemetry monitoring introduces an additional problem: a new prediction is produced at every timestep, and recomputing a full SWING explanation for every transition can be computationally redundant.

AdaptiveDeltaXAI extends Delta-XAI [1] with an adaptive control layer that determines when a fresh SWING explanation should be computed and when the most recent fresh map should instead be displayed as cached event context. Thus, SWING remains the underlying prediction-change attribution method, while AdaptiveDeltaXAI contributes the online refresh, detrending, staleness-control, temporal-display, and operational feature-group presentation mechanisms required for continuous monitoring.

Let $z_t$ denote the recomputation decision at timestep t. The decision variable is defined as

$$z_t \in \{0,1\}, \quad z_t = 1 \Leftrightarrow fresh\ Full - SWING\ call \quad (12)$$

A fresh explanation is computed only when the controller refreshes. Otherwise, the interface displays the latest fresh Full-SWING map [1] as cached_event_context, optionally attenuated for display by temporal discounting. Completeness is evaluated only for raw fresh Full-SWING maps; neither temporal discounting nor cache reuse is claimed to preserve completeness for the current transition. Let $a_t$ be the post-decision display age: $a_t = 0$ on a refresh and $a_t = a_{t-1} + 1$ on a cache hit. The displayed context $C_t$ is defined in Equation (13), so the first cache hit is displayed with age one rather than age zero.

$$C_t = \{E_t^{fresh}, z_t = 1, \gamma^{a_t} E_{t_{last}}^{fresh}, z_t = 0. \quad (13)$$

Here $t_{last}$ is the most recent refresh endpoint and $0 < \gamma \leq 1$. A fresh call sets $a_t = 0$. A cache hit increments the stored age before applying the display discount. The piecewise definition avoids referring to a nonexistent current fresh map on a cache-hit step. The map retains the observation-time coordinates of its originating refresh; reuse does not reassign its sensor-time entries to the current endpoint.

The controller evaluates a detector score $s_t$ rather than a per-sensor threshold. Before the current score updates the EMA, it forms the residual $r_t = |s_t - \mu_{t-1}|$; it then updates the score EMA $\mu_t$ and residual-scale EMA $\hat{\sigma}_t$. The controller quantities are

$$r_t = |s_t - \mu_{t-1}|, \mu_t = (1 - \alpha)\mu_{t-1} + \alpha s_t, \hat{\sigma}_t = (1 - \alpha)\hat{\sigma}_{t-1} + \alpha r_t. \quad (14)$$

The score state is initialized from the preceding score; the initial deviation is $max(|s_t - s_{t-1}|, \varepsilon)$, with $\varepsilon = 10^{-6}$. Refresh compares the selected trigger statistic with $kmax(\hat{\sigma}_{t-1}, \varepsilon)$ before any EMA update. When A2 is disabled, the numerator uses the raw adjacent change in Equation (15), but the saved code still updates the deviation EMA from the EMA score residual $r_t$ in Equation (14). Thus No-A2 is not a fully re-estimated raw-change threshold model. The residual scale is an EMA of absolute score residuals, not an estimate of their standard deviation.

$$r_t^{raw} = |s_t - s_{t-1}| \quad (15)$$

This score-level controller does not claim to remove every predictable telemetry cycle. Its effect is evaluated empirically in the three-seed ablation of Section 6.4.

These mechanisms distinguish AdaptiveDeltaXAI from Delta-XAI [1]. Full-SWING determines a fresh attribution when a prediction-change explanation is requested. AdaptiveDeltaXAI additionally determines when a fresh call is warranted, how cached event context expires, and how operational groups are displayed. It makes no claim that a cache hit is a fresh attribution of the current score change.

The framework consists of five components:

1. Adaptive triggering determines whether the current telemetry warrants a fresh SWING explanation.
2. EMA-based score-residual processing transforms the trigger signal; its effect on refresh behavior is evaluated empirically.
3. Input-drift and maximum-age invalidation refresh the cache when aggregate input change departs from its retained EMA state or the stored age reaches its limit.
4. Temporal discount weighting attenuates the displayed cached event context as it ages; it does not create or modify a fresh current-change explanation.
5. Operational feature-group presentation sums displayed channel scores within predefined temperature or power groups. In v1.9 this is a post-processing output, not a physical-topology model.

The method seeks to minimize amortized explanation latency while providing fresh explanations on refreshes and bounded-age cached event context otherwise.

Adaptive refresh trades computation against the age of displayed event context. Frequent refresh produces more current-transition explanations; cache reuse produces fewer. The following average cost is an operational objective for the heuristic controller, rather than an optimization problem solved with a measured run-time faithfulness constraint.

Over $T$ eligible decisions, define mean explanation-stage latency as

$$\underline{L} = \frac{1}{T}\sum_{t=1}^{T} \quad L_t \quad (cost\ to\ reduce) \quad (16)$$

The controller does not observe or enforce a run-time faithfulness threshold. Instead, raw fresh Full-SWING maps are evaluated offline through numerical completeness; cached event context receives no completeness or current-transition faithfulness claim.

$$\varepsilon_t^{fresh} = \left|\sum_{l\in H_t} \quad \sum_{d=1}^{D} \quad E_{t,l,d}^{fresh} - \Delta s_t\right| \quad (17)$$

With A3 enabled, cached event context is subject to the maximum-age safeguard. At the start of decision $t$, the controller tests the stored pre-decision age $a_{t-1}$ against $T_{max}$. If $a_{t-1} \geq T_{max}$, it refreshes and records $a_t = 0$; otherwise a cache hit increments and displays $a_t = a_{t-1} + 1$. Consequently, the displayed cache age is bounded by $T_{max}$, and the maximum-age refresh occurs on the subsequent decision. With no earlier trigger, a context last refreshed at $t_0$ can be displayed at ages 1 through $T_{max}$; the age-only refresh occurs at $t_0 + T_{max} + 1$.

$$0 \leq a_t \leq T_{max} \quad (18)$$

These definitions formalize the operational objective of AdaptiveDeltaXAI: reduce the cost of continuous fresh attribution while bounding the age of displayed cached event context. Fresh-map completeness is evaluated separately; it is not imposed on cache hits.

## 3. DATA & EXPERIMENTAL CONTEXT

### 3.1 The Cost of Unconditional Recomputation

Full-SWING [1] produces a fresh attribution map for each requested prediction change. In continuous telemetry monitoring, however, a new request can arise at every sampling interval, indefinitely, for as long as the system runs. The resulting systems constraint is the repeated cost of fresh Full-SWING calls, not an assumption that the original method is limited to isolated clinical or financial decisions.

In the repeated T4 workload (Section 6.3), the batched explanation stage costs 62.59 ± 6.49 ms per decision under always-refresh and 9.75 ± 0.24 ms under adaptive scheduling, a 6.42-fold ratio of reported means. This comparison excludes routine detector scoring and post-processing grouping. A cache hit returns event context, not a new prediction-change explanation.

Unconditional refresh incurs a fresh-attribution computation at every timestep. Whether observed latency meets a deployment target depends on hardware, detector, window size, integration resolution, batching, and competing system work. We therefore report the measured T4 distributions, the 2x2 batching x scheduling comparison, and the controlled 8-64-feature study rather than extrapolating a generic feasibility boundary.

Operational objective. Reduce the number of costly fresh calls while making cached displays explicit. Savings are reported for the measured explanation stage, with detector and presentation costs accounted for separately in the interpretation.

### 3.2 Periodicity Is Not Anomaly

A second challenge that is particularly salient in telemetry monitoring is strong, predictable periodic structure that a naive change trigger may not distinguish from genuine faults. Our controlled synthetic dataset encodes three periodic components intended to represent datacenter operating behavior: a diurnal ambient-temperature cycle, a short-period CRAC (computer room air conditioning) cycle, and a recurring batch-workload power cycle. Each produces measurable sensor variation without constituting a fault.

The ablation compares EMA-residual triggering with raw adjacent-score-change triggering. Removing A2 lowers the refresh rate in this configuration; this result does not by itself establish improved discrimination between periodic behavior and faults. Disabling A1 yields the defined always-refresh baseline. Matched-budget policy comparisons assess whether the adaptive placement of fresh calls improves event coverage.

## 3.3 Faults Do Not Arrive in One Shape

We evaluate three controlled fault patterns relevant to telemetry monitoring: isolated bursts, overlapping bursts, and gradual drifts. These patterns test abrupt, concurrent, and progressive changes in the model score; they are controlled synthetic conditions, not observed production-failure distributions.

The three controlled regimes expose different score dynamics. Isolated bursts test localized transitions; overlapping bursts test concurrent changes; and gradual drifts test progressive changes. The evaluation reports aggregate results and does not rank these regimes by intrinsic difficulty without regime-specific evidence.

## 3.4 Operational Grouping of Heterogeneous Sensors

The present implementation uses predefined operational feature groups as a presentation proxy, rather than a measured physical rack, cooling-zone, or PDU graph. Grouping temperature features or power input/output features provides a compact operational display, but it should not be interpreted as validated physical topology.

A5 sums the selected displayed map into predefined operational groups. The frozen ablation passes a use_i5 flag, but run_adaptive does not consume that flag and grouping is generated separately. No-A5 therefore repeats the same controller computation: its identical refresh rate and map metrics do not test whether grouping improves interpretation, and timing differences between these runs are not an A5 effect.

# 4. ADAPTIVEDELTAXAI: A FRAMEWORK FOR CONTINUOUS TELEMETRY EXPLANATION

## 4.1 SWING Baseline

Our fresh-attribution implementation follows the Full-SWING near/far shifted-window construction in Eq. (9) and Appendix C. It passes internal structural and linear-completeness tests. We do not claim that the prior adjacent window is the only meaningful baseline, nor do we claim direct numerical parity with an external Delta-XAI implementation. For fresh maps, numerical completeness is measured rather than asserted as exact.

- Fresh-map numerical completeness: the sum of raw fresh attributions approximately reconstructs the observed prediction change. Residuals are reported explicitly in the synthetic three-seed results. Cached or temporally discounted displays have no completeness claim.
- Implementation invariance is a theoretical attribution property. It is not separately evaluated as an empirical validation gate in this study.
- Skew symmetry is a property discussed for prediction-difference attribution. It is not separately reported as an empirical validation gate in the experiments.

## 4.2 Five Domain Specific Adaptations

The five components have different roles. A1 and A2 govern the score trigger; A3 supplies drift and age safeguards; A4 discounts cached displays; A5 groups the selected output after the controller. A1–A4 are exercised in the decision/display loop. The saved No-A5 flag has no effect in that loop, so the corresponding row is an implementation control rather than a validated grouping ablation.

We measure refresh and numerical-completeness behavior for the controller variants, and treat A4/A5 presentation utility as unevaluated. Temporal discounting and grouping do not create fresh attribution information.

### A1: Adaptive Score-Residual Triggering

AdaptiveDeltaXAI maintains cached_event_context, its age $a_t$, a score EMA, a residual-scale EMA, and input-change EMA state. At each online step it computes the detector score and uses only the current observation, current score, preceding score, and previously retained controller state. An empty cache always produces the initialization refresh.

$$z_t = 1\{empty \lor r_t > kmax(\hat{\sigma}_{t-1}, \varepsilon) \lor d_t > k_{drift} \lor a_{t-1} \geq T_{max}\} \quad (19)$$

For normalized observations, define $q_t = D^{-1} \sum_{d=1}^{D} |x_{t,d} - x_{t-1,d}|$ and $d_t = |q_t - \bar{q}_{t-1}|/max(v_{t-1}, \varepsilon)$. After the refresh decision, update $\bar{q}_t = (1-\alpha)\bar{q}_{t-1} + \alpha q_t$ and $v_t = (1-\alpha)v_{t-1} + \alpha|q_t - \bar{q}_{t-1}|$. At the first decision, initialize $\bar{q}_{t-1} = 0$ and $v_{t-1} = \varepsilon$; the empty cache forces a fresh call. Equation (19) describes the full configuration. No-A1 explicitly forces always-refresh; No-A3 removes both the drift and age conditions.

**A2: EMA-Based Score Detrending**

$$r_t^{A2} = |s_t - \mu_{t-1}|, \qquad r_t^{No\ A2} = |s_t - s_{t-1}| \quad (20)$$

A2 uses the EMA score residual in Eq. (20) rather than raw adjacent score change. The three-seed ablation shows that A1, not A2, is the component whose removal forces 100% refresh. Removing A2 changes the operating point and yields a lower measured refresh rate in this configuration; no universal necessity claim is made for A2.

**A3: Input-Drift and Maximum-Age Invalidation**

A3 forces refresh when standardized aggregate input-change drift exceeds its configured threshold or when the pre-decision cache age $a_{t-1}$ reaches $T_{max}$. It is therefore both a live-signal safeguard and a maximum-age bound. It limits the persistence of cached event context; it does not make a non-refresh display a current-transition explanation.

**A4: Display-Only Temporal Weighting**

$$C_t = \gamma^{a_t} E_{t_{last}}^{fresh} \quad on\ cache\ hits\ only \quad (21)$$

A4 applies Equation (21) only after a cache hit. Its temporal attenuation is a display convention; no operator study establishes that attenuation alone communicates age reliably. The saved records explicitly include age and output type. Raw fresh maps are not discounted.

**A5: Operational Feature-Group Presentation**

For temperature, A5 groups features 0–3 and 4–7 as thermal_zone_A and thermal_zone_B. For power, it groups features 0–1 as PDU_input and feature 2 as PDU_output. Displayed scores are summed over time, then within each group. These labels denote fixed index groups, not measured physical zones or inferred fault propagation. Grouping is generated after run_adaptive and lies outside its step timer.

## 4.3 Full Trigger Logic and Amortized Cost

A fresh SWING computation [1] is triggered when the cache is empty, when the pre-update score-residual condition fires, or when A3 detects input-change drift or a pre-decision maximum-age condition. Otherwise, the system presents cached_event_context. The non-refresh branch avoids a Full-SWING call but does not create a new current-change attribution.

$$\underline{L} = \rho \underline{L}_{refresh} + (1 - \rho)\underline{L}_{hit} \quad (22)$$

Here $\rho$ is the refresh fraction over the declared workload. The barred branch latencies are measured means on refresh and cache-hit decisions, so Equation (22) is the empirical mixture identity for that workload. A fresh-call timer measures the attribution computation separately; the step timer measures the explanation-control loop described in Section 2.3. It excludes the precomputed detector scores and subsequent operational grouping. Rates and timings from different workloads must not be substituted into the same mixture.

Figure 2 shows the decision loop and post-processing. The score residual and input-drift statistics are formed from retained pre-update state. The decision selects either a fresh Full-SWING map or cached event context; the latter alone receives A4 discounting. A5 groups the resulting display after the timed controller loop.

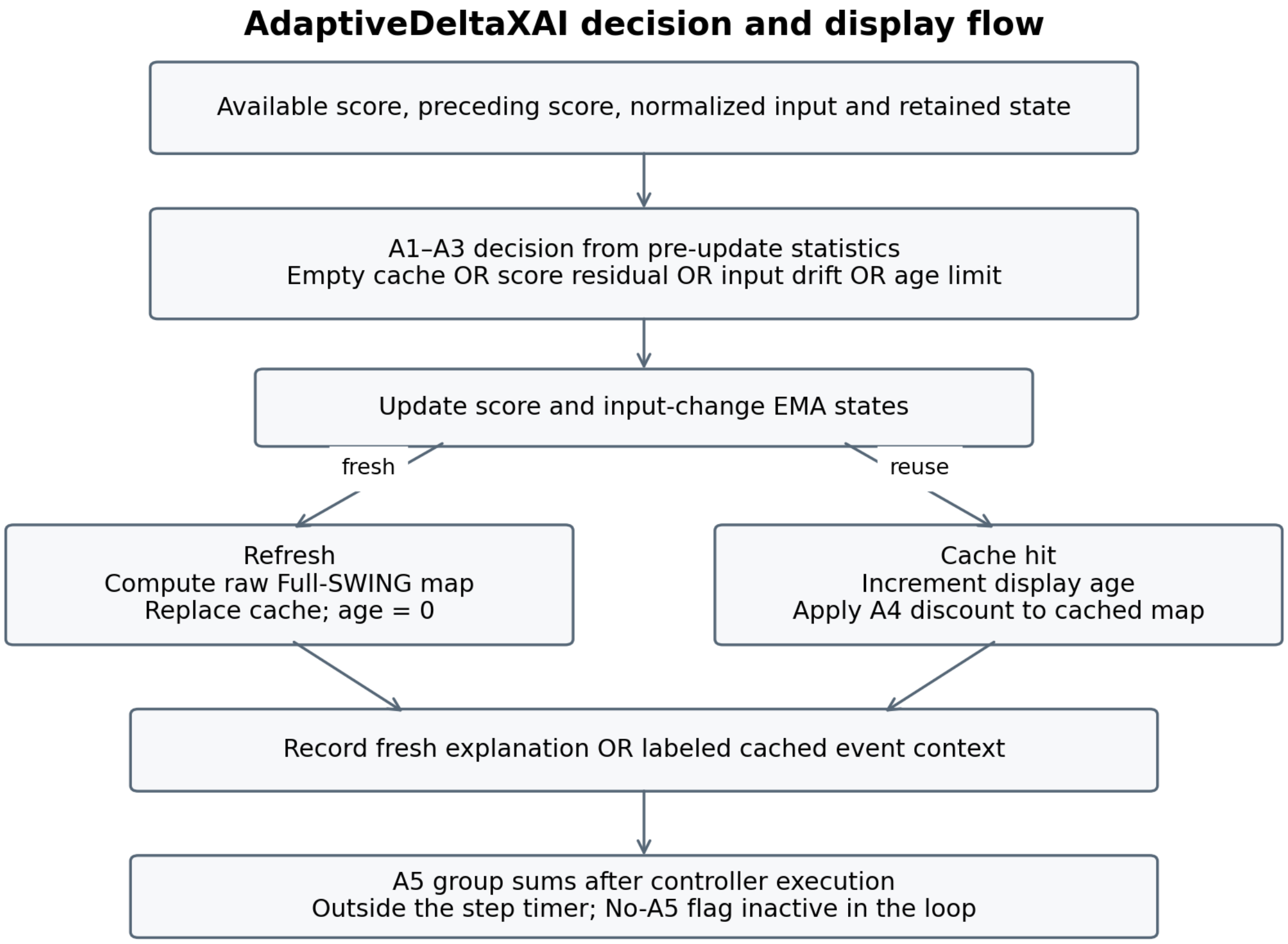

Figure 2: Adaptive decision loop and output semantics. A1–A3 select fresh computation; A4 discounts only cache hits. EMA state updates follow the decision. A5 is separate post-processing, outside the controller timer. Only raw fresh maps receive a current-transition completeness assessment.

# 5. EVALUATION METRICS FOR EXPLAINING PREDICTION CHANGES

## 5.1 Detector Quality Metrics

We report AUROC and average precision (AP, labeled AUPRC in the frozen files) for threshold-free detection. Thresholded synthetic metrics use $p \geq 0.5$ and include precision, recall, F1, MCC, balanced accuracy, specificity, and event detection. An event is a contiguous positive-label interval; its detection delay is measured to the first thresholded positive and is conditional on detection. Event detection and explanation-refresh coverage use different populations (Section 6.2) and must be read separately.

## 5.2 Current-Score Perturbation Proxy

$$AOPC_s(X;K) = \frac{1}{K}\sum_{k=1}^{K} \quad [f(X) - f(M_k X)] \quad (23)$$

Equation (23) defines our time-series adaptation of perturbation-based attribution evaluation [21]. The cited work motivates the general region-perturbation approach; the sensor-lag ranking, training-normalized zero replacement, and $K = 8$ masking protocol below are our experimental choices, rather than a reproduction of its original image-region perturbation procedure. $M_k$ replaces the top $k$ sensor-lag cells, ranked by absolute attribution, with the training-normalized zero reference; $K = 8$. The same 50 positive held-out temperature windows are used for all methods at seed 42. Full-SWING uses a prefix-padded history and its current-window slice, while AW-IG uses a constructed preceding window formed by repeating the first observation. This protocol does not explain or perturb a real adjacent-window score difference. Individual method intervals use window bootstrap resampling; the paired Full-SWING-minus-comparator intervals use 2,000 event-cluster bootstrap resamples across 11 clusters. The two interval types are distinct.

KernelSHAP and LIME backgrounds in the supplementary current-score proxy are sampled from held-out inputs, without using their labels for background selection. Fresh-method timing uses the first 30 held-out windows as background. These are not training-only background protocols. The direct prediction-difference experiment instead uses 30 training-partition unions. KernelSHAP uses 100 samples; LIME leaves its perturbation-sample count at the library default rather than sharing the 100-sample budget. Explainer setup is outside the reported method timing, so comparisons describe these configured implementations rather than equal computational budgets.

Notation and selected frozen configuration.

| Symbol | Meaning | Setting / unit |
|---|---|---|
| $E_t$; $e_t$ | Raw fresh sensor-time map; sum over absolute time per channel | $(W+1) \times D$; $D$ |
| $H_t$; $J_t$; $J_{t-1}$ | Union coordinates; current/previous embedding operators | $\{t-W, \ldots, t\}$ |
| $\rho$ | Measured fresh-call fraction for a declared workload | Unitless |
| $q_t$; $d_t$ | Mean adjacent normalized input change; standardized drift | Drift threshold 5.0 |
| $a_t$ | Post-decision display age | 0 on fresh; first hit 1 |
| $W$; $L$; $\delta$ | Window; retrospective lag; endpoint separation | 20; 5; 1 observations |
| $K$ | Maximum cumulative perturbation mask level | 8 cells |

## 5.3 Efficiency: Recompute Rate and Amortized Latency

We report refresh rates, fresh-call latency, and amortized explanation-stage latency. The five-method latency distribution contains five timed repetitions on each of 10 windows, rather than 50 independent windows. The separate batching × scheduling study contains five repetitions per condition with controller state reset each time. Warm-up runs are excluded; initialization decisions are included.

## 5.4 Component Ablation Methodology

We compare Full with No-A1 through No-A5 for seeds 42, 52, and 62. Each temperature ablation initializes state on a test-tail input starting 26 observations before the test boundary; its 46-observation controller guard places the first endpoint at absolute index 13,843 and target at 13,844. Each run has 3,437 operational decisions through endpoint 17,279, including the final invalid target 17,280. The reported latency, refresh, and fresh-completeness summaries include that final operational decision. The AOPC proxy uses only the first 50 fresh maps selected by each configuration and is not a common-window attribution-quality comparison. No-A5 is an inactive-flag control, as explained in Section 3.4.

# 6. EXPERIMENTS

All numerical experiments reported here use the frozen Master v1.9 artifacts. Synthetic detector, full-stream controller, ablation, scaling, and architecture studies use seeds 42, 52, and 62. Method comparisons use seed 42 and 50 common positive held-out targets. Public results cover SMD machine-1-1, machine-1-2, machine-1-3, and PSM, with 50 selected fresh maps per unit. Timings were recorded on a Tesla T4 after the stated warm-ups. Appendix D identifies the different workloads and state policies; Appendix F records what the reconciliation can establish from saved artifacts.

## 6.1 Experimental Setup

Features and target. The synthetic design provides eight temperature channels: Inlet_Temp, CPU1_Temp, CPU2_Temp, DIMM_A1_Temp, DIMM_B1_Temp, PCH_Temp, Exhaust_Temp, and Fan1_norm. The power channels are PS1_Input_W, PS2_Input_W, and PS1_Output_W. The specified sampling interval is 10 seconds over 48 hours. Synthetic classifiers use a 20-observation window and predict the next observation's binary anomaly label. Held-out positive prevalence is $331/3436 = 9.63\%$. The full-stream fault design is summarized in Appendix A.

Detector models. Temperature and power use independent two-layer GRUs with 64 hidden units, a 32-unit ReLU head, 0.20 inter-layer dropout, and 0.10 head dropout. Training minimizes class-weighted binary cross-entropy. The Transformer variant is specified in Section 6.7. Public experiments use a distinct GRU final-observation reconstruction model with a log1p-MSE score (Section 6.6).

Training and evaluation. For the synthetic classifiers, we use five epochs, batch size 256, AdamW learning rate $5 \times 10^{-4}$, and weight decay $10^{-4}$. The checkpoint with the lowest validation loss is retained. Normalization uses training observations only; detector windows stay within each split. Model weights remain fixed within each attribution/policy comparison. Seeds change training initialization; feature scaling and architecture experiments retrain their detectors. The frozen configuration does not document a separate hyperparameter search or establish that all design choices were selected before inspecting this dataset.

Explanation interface. A differentiable score function accepts a $(1, W, D)$ tensor and returns a scalar. The controller makes decisions from the current score, preceding score, current normalized observation, and retained state. The supplied runner precomputes the score sequence before the loop for execution, but its adaptive decision at each endpoint reads no later score. That precomputation is excluded from the step timer.

Figures 3 and 4 reproduce the seed-42 temperature and power detector ROC and precision–recall curves from the frozen v1.9 outputs. They illustrate one seed; Table 1 summarizes all three seeds.

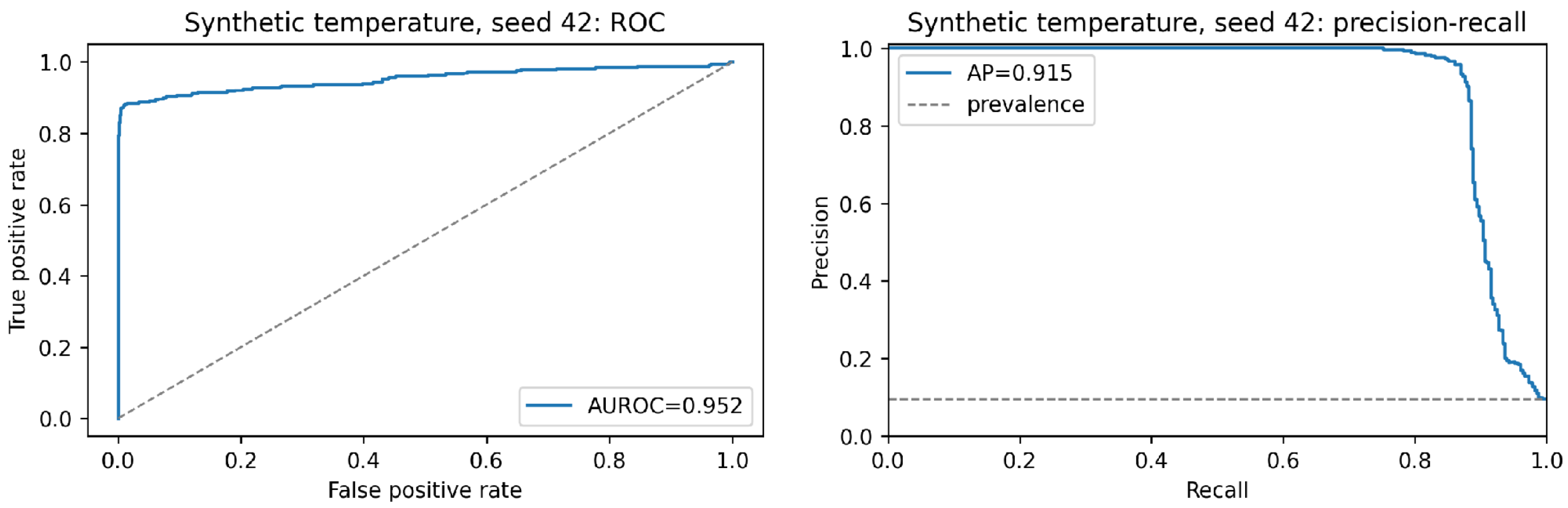

Figure 3: Held-out temperature detector ROC and precision–recall curves, seed 42. Source: v1.9 synthetic_temperature_seed42_roc.png and synthetic_temperature_seed42_pr.png. AUPRC denotes average precision.

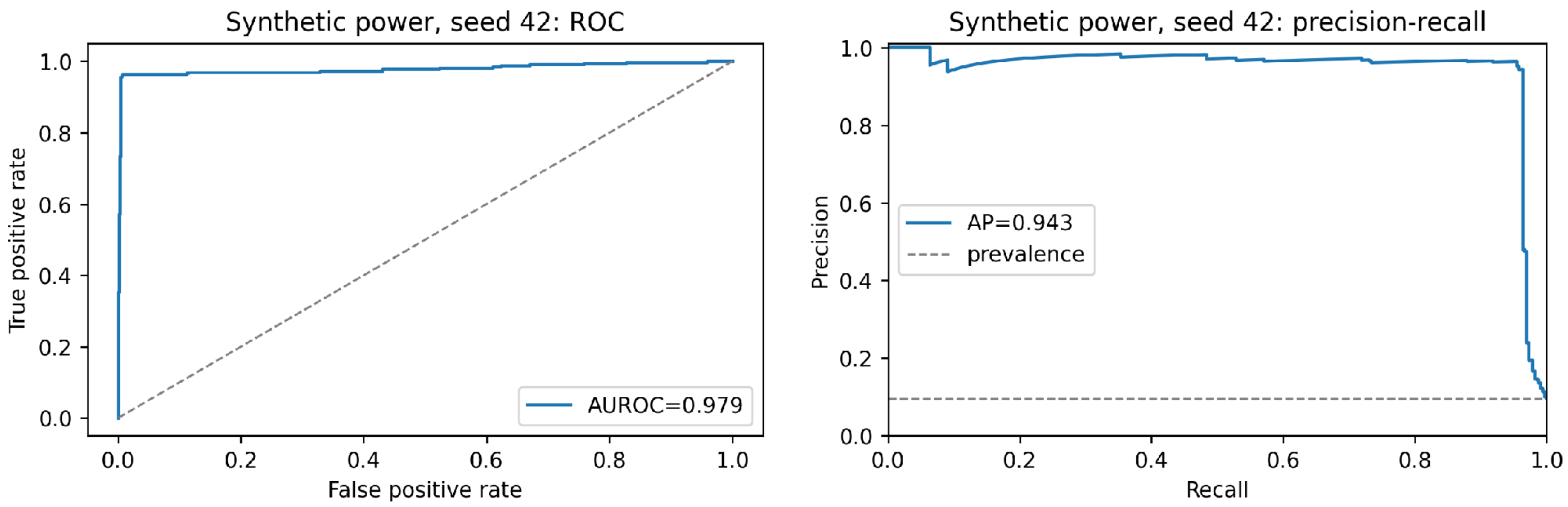

Figure 4: Held-out power detector ROC and precision–recall curves, seed 42. Source: v1.9 synthetic_power_seed42_roc.png and synthetic_power_seed42_pr.png.

## 6.2 Detector Quality

Table 1 reports mean ± sample standard deviation across three seeds. The split-local detector set contains 3,436 windows, 331 positive labels, and 12 positive intervals per seed. Continuous-controller refresh coverage in Table 6 instead includes all 3,456 held-out target indices and 13 intervals. Detector event recall of 100% and refresh coverage of 92.3% therefore describe different actions and populations.

Table 1: Three-seed synthetic held-out detector metrics. AUROC/AUPRC are threshold-free; the remaining values use the stated fixed 0.5 classifier threshold. Values are mean ± standard deviation.

| Metric | Temperature | Power |
|---|---|---|
| AUROC | 0.950 ± 0.005 | 0.979 ± 0.001 |
| Average precision (AUPRC) | 0.913 ± 0.006 | 0.945 ± 0.007 |
| MCC | 0.895 ± 0.010 | 0.943 ± 0.005 |
| Balanced accuracy | 0.924 ± 0.009 | 0.976 ± 0.002 |
| Specificity | 0.996 ± 0.001 | 0.993 ± 0.001 |
| False-positive rate | 0.004 ± 0.001 | 0.007 ± 0.001 |
| Precision | 0.960 ± 0.013 | 0.937 ± 0.007 |
| Recall | 0.852 ± 0.018 | 0.960 ± 0.003 |
| F1 | 0.903 ± 0.010 | 0.948 ± 0.004 |
| Detector event recall (detected / total) | 100% (12 / 12) | 100% (12 / 12) |
| Detection delay, detected only (steps) | 3.97 ± 0.43 | 1.11 ± 0.10 |
| False-alarm episodes / 1,000 normal steps | 0.000 ± 0.000 | 0.107 ± 0.186 |
| Detector events missed | 0 | 0 |

## 6.3 Five Method Benchmark: Measured Latency

Table 2 reports five fresh-method implementations measured on T4. Each has 50 trials: five timed repetitions on each of 10 seed-42 positive held-out windows, with one warm-up per window. Full-SWING timing uses prefix-padded history, rather than the real history used by the operational controller and direct difference experiment. KernelSHAP uses 100 samples; LIME uses its default perturbation count, and explainer construction is excluded. These results are configuration-specific costs, not a controlled comparison at equal perturbation budgets.

Table 2: *T4 fresh-attribution latency distribution across 50 trials per method. Mean, median, P95, P99 are measured directly.*

| Method | Mean ms | Median / P95 / P99 ms |
|---|---|---|
| AW-IG | 12.13 | 11.80 / 15.80 / 16.86 |
| Full-SWING [1] | 59.27 | 57.20 / 71.27 / 76.72 |
| KernelSHAP [5] | 31.18 | 30.40 / 35.18 / 42.38 |
| LIME [4] | 283.26 | 251.96 / 558.47 / 667.54 |
| Zero-baseline IG [6] | 13.37 | 12.89 / 14.67 / 23.16 |

Table 3 isolates batching from scheduling. The probe contains 146 observations and the controller begins at local endpoint 45, yielding 101 decisions per repetition. Adaptive runs refresh $17/101 = 16.83\%$ of decisions. The batched means imply an explanation-stage cost ratio of $(62.59\,ms)/(9.75\,ms) = 6.42$. The 101-decision probe and its initialization are distinct from the 17,235-decision full-stream workload, whose temperature refresh rate is 8.09%. Routine detector inference and A5 grouping are outside these timing summaries.

Table 3: T4 batching × scheduling study, mean ± sample standard deviation over five repetitions. Each repetition contains 101 operational decisions, including initialization; adaptive refreshes 17. Detector-score precomputation and A5 grouping are excluded.

| Condition | Refresh % | Amortized ms | Fresh-call ms |
|---|---|---|---|
| batched always refresh | 100.00 ± 0.00 | 62.59 ± 6.49 | 62.35 ± 6.47 |
| batched adaptive refresh | 16.83 ± 0.00 | 9.75 ± 0.24 | 57.45 ± 1.38 |
| unbatched always refresh | 100.00 ± 0.00 | 2626.08 ± 38.40 | 2625.78 ± 38.40 |
| unbatched adaptive refresh | 16.83 ± 0.00 | 443.52 ± 5.49 | 2634.50 ± 32.65 |

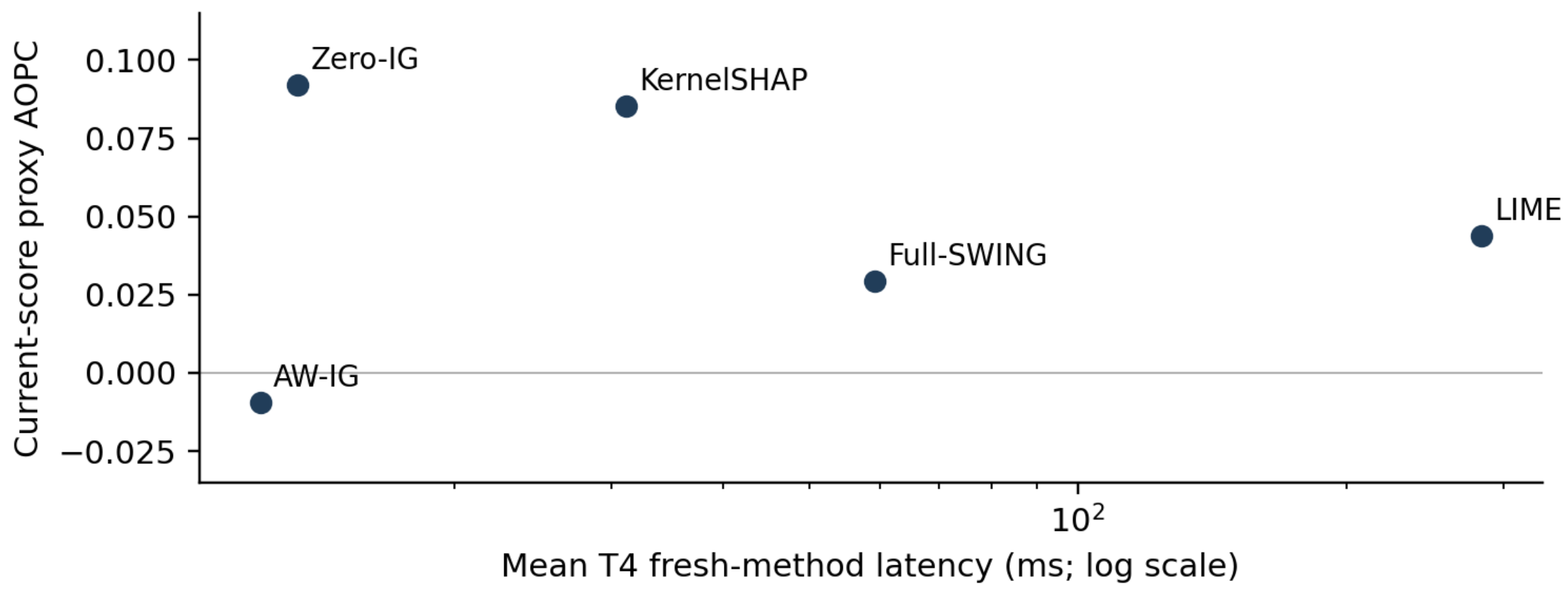

Figure 5: v1.9 current-score mean-masking AOPC proxy versus measured mean T4 fresh-attribution latency. This 50-window, seed-42 proxy does not establish a universal method winner or directly assess prediction-difference faithfulness.

Table 4 reports the supplementary current-score proxy. The direct prediction-difference experiment is reported separately in Table 16 and Equation (24), with actual observed history, common target indices, and a training-only background for KernelSHAP/LIME. It is the more direct test of the stated prediction-change objective; current-score proxy values and difference-perturbation values must not be pooled.

Direct prediction-difference protocol. For each of 50 seed-42 positive held-out temperature targets, all methods rank magnitudes on the same $21 \times 8$ absolute-time union. At each $k = 1, \ldots, 8$, $M_k$ masks the same selected absolute sensor-time coordinates to normalized zero wherever they appear in the preceding and current windows. The metric averages the reduction in absolute score difference. Positive values indicate a reduction; negative values mean masking increased that difference. KernelSHAP and LIME explain the union-to-score-difference function using training-union backgrounds. Zero-IG subtracts separately aligned previous/current zero-baseline maps. AW-IG is computed between adjacent windows and embedded at current-window positions, as implemented; it does not use the Full-SWING multipath alignment construction.

The mean direct scores are −0.0154 for Full-SWING, 0.0061 for AW-IG, 0.0236 for Zero-IG, 0.0261 for KernelSHAP, and 0.0106 for LIME. Full-SWING-minus-comparator exploratory 95% intervals are below zero for Zero-IG, KernelSHAP, and LIME and include zero for AW-IG (Table 16). Thus these data do not support Full-SWING superiority under this perturbation test. The window bootstrap does not account for event dependence or multiple comparisons; this single-seed, temperature-only result does not establish a universal method ranking.

Table 4: Supplementary current-score mean-masking proxy, 50 common windows. Method intervals use window bootstrap; paired Full-SWING-minus-method intervals use event-cluster bootstrap across 11 clusters. Both use 2,000 resamples. These are different interval constructions.

| Method | Proxy mean | Window 95% CI | Paired difference [cluster 95% CI] |
|---|---|---|---|
| AW-IG | -0.0095 | [-0.0562, 0.0316] | +0.0387 [0.0069, 0.0722] |
| Full-SWING [1] | 0.0292 | [-0.0112, 0.0731] | Reference |
| KernelSHAP [5] | 0.0852 | [0.0273, 0.1414] | -0.0560 [-0.1245, 0.0166] |
| LIME [4] | 0.0437 | [-0.0268, 0.1144] | -0.0145 [-0.1062, 0.1034] |
| Zero-baseline IG [6] | 0.0920 | [0.0127, 0.1738] | -0.0628 [-0.1434, 0.0447] |

## 6.4 Component Ablation Study

Table 5 reports the three-seed six-configuration ablation on the synthetic temperature stream, following the common protocol in Section 5.4.

Table 5: Temperature controller variants, mean ± sample standard deviation over three seeds and 3,437 operational decisions per run. AOPC uses the first 50 configuration-selected fresh maps. No-A5 is an inactive-flag control rather than a grouping-effect ablation.

| Variant | Refresh % | Amort. ms | Fresh ms | Fresh residual | Proxy AOPC |
|---|---|---|---|---|---|
| Full | 8.20 ± 0.09 | 5.41 ± 0.03 | 65.20 ± 0.93 | 0.000199 ± 0.000029 | 0.0173 ± 0.0595 |
| No A1 | 100.00 ± 0.00 | 66.21 ± 1.69 | 65.96 ± 1.69 | 0.000055 ± 0.000008 | -0.0182 ± 0.0075 |
| No A2 | 2.90 ± 0.26 | 1.94 ± 0.14 | 65.38 ± 1.29 | 0.000077 ± 0.000006 | -0.0451 ± 0.0335 |
| No A3 | 7.26 ± 0.09 | 4.78 ± 0.01 | 65.08 ± 0.89 | 0.000221 ± 0.000032 | 0.0302 ± 0.0605 |
| No A4 | 8.20 ± 0.09 | 5.25 ± 0.12 | 63.35 ± 1.77 | 0.000199 ± 0.000029 | 0.0173 ± 0.0595 |
| No A5 | 8.20 ± 0.09 | 5.36 ± 0.26 | 64.64 ± 3.23 | 0.000199 ± 0.000029 | 0.0173 ± 0.0595 |

No-A1 is the explicitly defined always-refresh reference; its 100% rate is a configuration outcome. No-A2 lowers the measured refresh rate while retaining the EMA-residual deviation update. No-A3 removes drift and age refresh conditions. No-A4 changes only cached displays. No-A5 does not change the executed controller because its flag is unused; differences in its measured latency are run variability and cannot be attributed to grouping. Public matched-budget controls in Section 6.6 provide separate evidence about refresh placement.

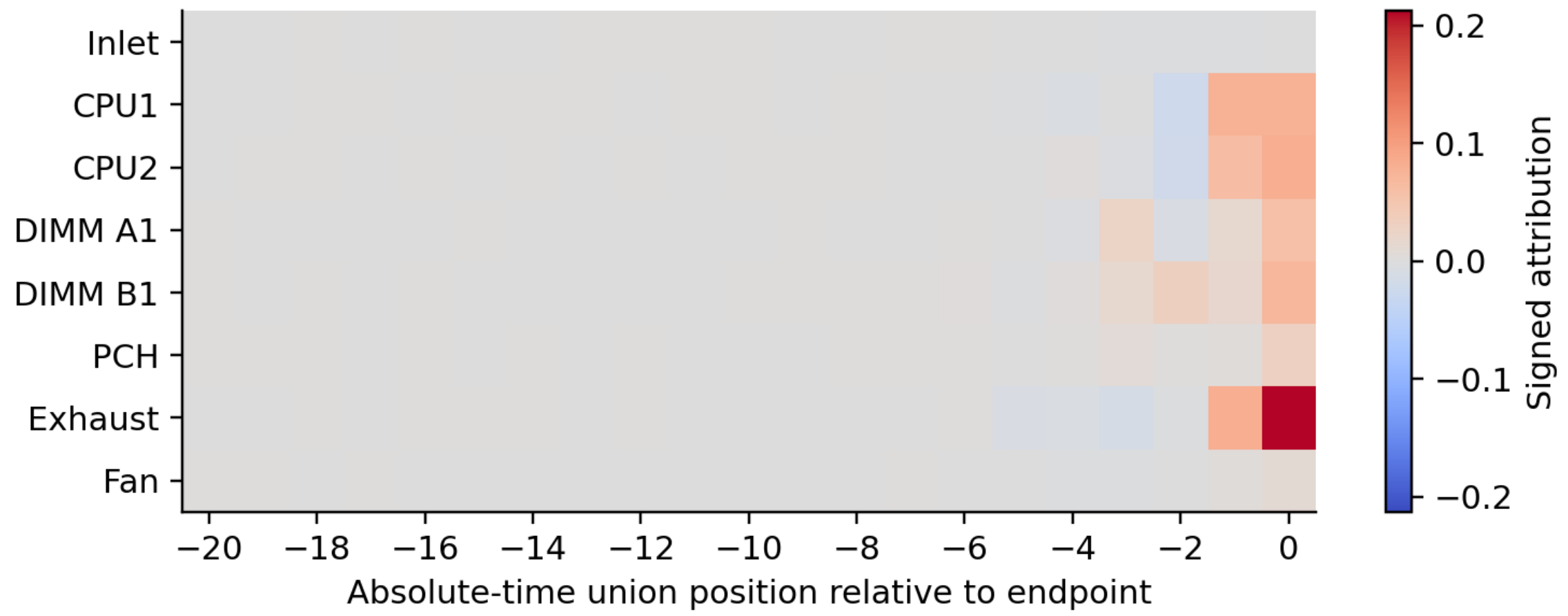

Figure 6: Raw fresh Full-SWING map at endpoint 16075 and target 16076, seed 42. The full 21-position adjacent-window union is shown with a zero-centered color scale; $|\Delta s_t| = 0.783$. Completeness concerns the sum of this full map, not a cached display.

All ablation configurations have three saved seed runs. Their mean and standard deviation describe the declared operational interval and the first-50-fresh-map proxy subset; they do not imply a paired attribution comparison on identical maps.

AOPC is a current-score mean-masking proxy evaluated on fresh attribution maps selected by each configuration's own controller. The values therefore describe those selected fresh maps and do not isolate a component's effect on attribution quality.

Figure 6 provides a representative raw fresh Full-SWING sensor-lag map, while Figure 1 and Table 6 separately report when refreshes occur, held-out-label explanation-refresh coverage, missed events, and first-refresh delay. For Table 6, labels are filtered to the held-out interval from controller records generated during a continuous full-stream pass; EMA and cache state therefore carry into the held-out interval. An event is refresh-covered when a contiguous labeled positive interval contains at least one fresh explanation; delay is measured from interval onset to the first fresh refresh and is conditional on coverage.

Table 6: *Held-out synthetic explanation-refresh coverage. Values are mean ± standard deviation across three seeds; delay is conditional on covered contiguous positive-label intervals. Held-out labels are evaluated under a continuous full-stream controller state; EMA and cache state carry into the held-out interval.*

| Stream | Events | Fresh-explanation event coverage | Missed refresh events | Delay steps | Delay seconds | Fresh / targets |
|---|---|---|---|---|---|---|
| Temperature | 13 | 12/13 (92.3%) | 1 | 3.42 ± 0.17 | 34.2 ± 1.7 | 264.7 ± 2.5 / 3,456 |
| Power | 13 | 12/13 (92.3%) | 1 | 0.89 ± 0.05 | 8.9 ± 0.5 | 151.0 ± 5.2 / 3,456 |

## 6.5 Sensor Count Scaling Study

The scaling study tiles the eight temperature channels to 8, 16, 32, or 64 features and adds independent Gaussian noise with standard deviation $10^{-3}$. A new GRU is trained for each scale and seed. The same test-tail controller interval as the ablation gives 3,437 operational decisions per run; normalization remains training-only. The features are correlated replicas, rather than independent added physical sensors.

Table 7 reports three-seed controlled scaling. Across the tested 8–64 feature configurations, fresh-call latency was approximately 67.5–68.3 ms and amortized latency approximately 5.45–5.71 ms. These measurements describe this workload and do not establish a general scaling law.

Table 7: Three-seed controlled feature-count scaling. Values are mean ± standard deviation on the Tesla T4.

| Features | Refresh % | Amortized ms | Fresh ms | Fresh residual |
|---|---|---|---|---|
| 8 | 8.18 ± 0.08 | 5.65 ± 0.37 | 68.30 ± 3.91 | 0.00020 ± 0.00003 |
| 16 | 8.31 ± 0.04 | 5.71 ± 0.25 | 67.96 ± 3.30 | 0.00028 ± 0.00009 |
| 32 | 7.88 ± 0.14 | 5.45 ± 0.36 | 68.34 ± 3.58 | 0.00028 ± 0.00004 |
| 64 | 7.95 ± 0.24 | 5.45 ± 0.21 | 67.52 ± 0.71 | 0.00050 ± 0.00005 |

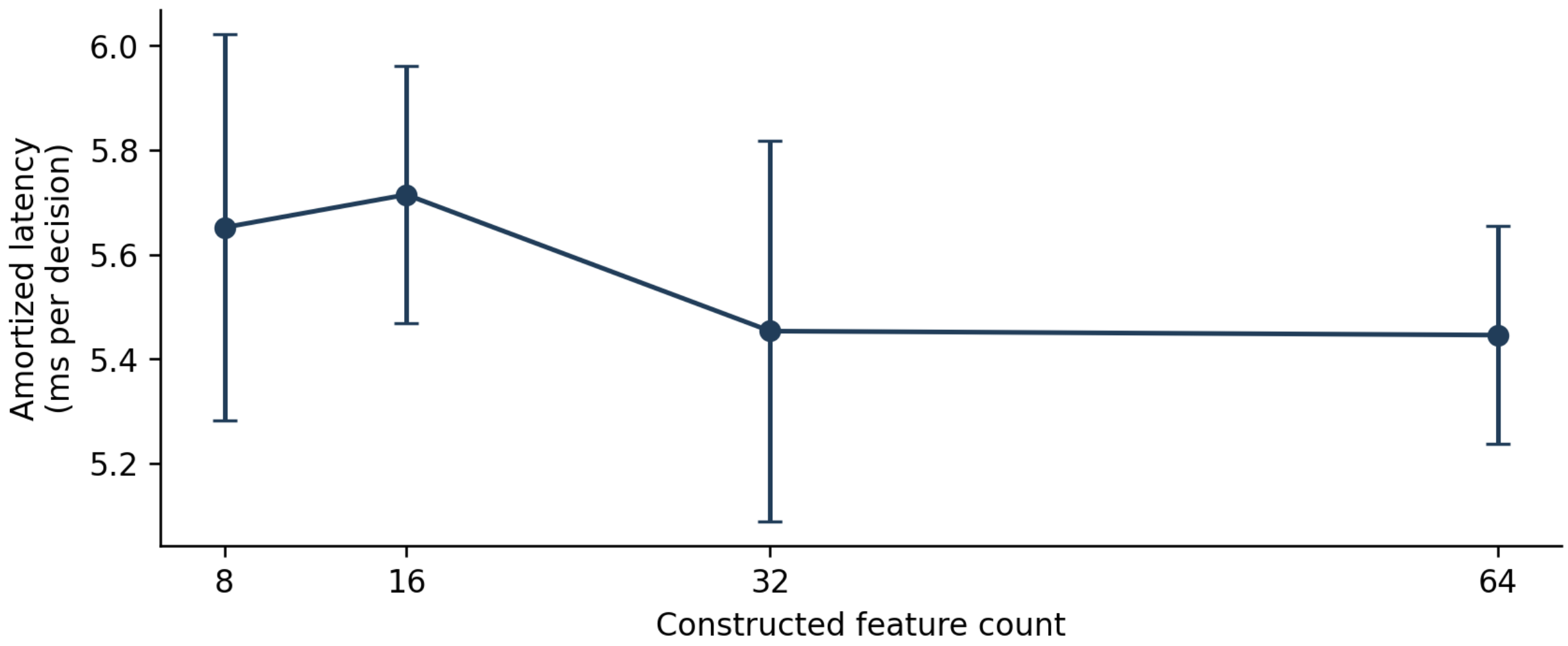

Figure 7: v1.9 three-seed controlled feature-count scaling on T4. Error bars show standard deviations; measurements are not extrapolated beyond the tested synthetic configurations.

Across 8 to 64 features, the observed fresh-call and amortized latencies are nearly flat under the tested batched implementation. This is a measured result for this controlled setup, not a general scaling law. The experiment does not establish super-linear growth and does not justify extrapolation beyond the measured range.

Because expansion replicates channels with small noise and changes the trained model and schedule, the nearly flat timing curves establish behavior only for these correlated synthetic workloads. They do not establish performance on 64 distinct physical sensors or a production topology.

Scaling completeness is reported for raw fresh maps selected by each resulting controller. It is not a controlled comparison on a fixed set of score transitions across dimensions.

## 6.6 Public Dataset Validation

We evaluate a documented public subset: SMD [2] machines machine-1-1, machine-1-2, and machine-1-3, plus PSM [3]. A GRU-based reconstruction detector inspired by recurrent encoder-decoder anomaly detection [20] is trained on the public training partition and produces log1p reconstruction-MSE scores. Each end-exclusive reconstruction window $x[t-W:t]$ is aligned to the label of its final reconstructed observation, $y[t-1]$. We report full-test threshold-free AUROC/AUPRC as the primary public detection evaluation, exact matched-budget policy controls in score-only scheduling mode, and 50 fresh Full-SWING maps [1] per unit. The public study does not claim all-28-machine coverage or broad public faithfulness superiority. The 50 public Full-SWING maps per unit are selected uniformly without replacement from aligned positive windows with seed 42.

The public score-only scheduler uses $\alpha = 0.05$, $k = 2.5$, $T_{max} = 60$, and no input-drift trigger. Its initial EMA is the first score and deviation starts at $10^{-6}$. The implementation forces the first mask entry true after generating the mask; this does not reset the age counter during the mask-construction loop. The public scheduler is therefore a distinct scheduling protocol, rather than a full A1–A5 controller run. Retrospective per-unit budgets are used for all controls.

Table 8: *Public full-test threshold-free detection metrics with reconstruction-label alignment to the final observation in each end-exclusive window. The public scope is three named SMD units [2] plus PSM [3].*

| Dataset / unit | Windows | Anomaly % | AUROC | AP | AP lift |
|---|---|---|---|---|---|
| SMD machine-1-1 | 28,459 | 9.47 | 0.877 | 0.534 | 5.64× |
| SMD machine-1-2 | 23,674 | 2.29 | 0.868 | 0.221 | 9.63× |
| SMD machine-1-3 | 23,683 | 3.45 | 0.788 | 0.201 | 5.83× |
| PSM | 87,821 | 27.76 | 0.771 | 0.570 | 2.05× |

Across the three SMD units, adaptive refresh rates are 5.77%, 5.91%, and 5.86%; PSM refreshes 5.50% of test windows. Periodic and random controls use the realized adaptive refresh count per unit. The random control uses one fixed seed-42 schedule; the score-change comparator uses the completed score sequence and is an offline oracle. These scheduling results cover only the documented public subset.

Table 9: *Retrospectively budget-matched public policy controls. Periodic and random schedules receive the adaptive policy's realized per-unit refresh count; once count and horizon are fixed, they do not use future scores. The score-change row is an offline oracle.*

| Policy | SMD macro coverage | PSM coverage |
|---|---|---|
| AdaptiveDeltaXAI | 100.0% | 75.0% |
| Periodic, matched count | 82.8% | 52.8% |
| Random, matched count | 86.1% | 40.3% |
| Offline score-change oracle | 100.0% | 81.9% |

Public detector quality ranges from AUROC 0.788 to 0.877 across the three SMD units; PSM yields AUROC 0.771 and AP 0.570. Table 17 gives fresh-only numerical-completeness residuals for 50 selected positive windows per unit, all below the configured mean/P95 gates of 0.05/0.15. Public scheduling is score-only; the separate subset maps do not measure continuous public attribution latency or comparative faithfulness. Supplementary chronological calibration uses a separately fitted detector and is reported in Appendix G because its thresholded performance is uneven. The supplied public reconstruction training routine uses AdamW with learning rate $10^{-3}$, five epochs, and batch size 256. It does not explicitly override weight decay and retains the final training epoch rather than selecting a checkpoint by validation loss.

## 6.7 Two-Architecture Validation

We compare the temperature GRU with a two-layer Transformer encoder using the same chronological split, training settings, seeds, and controller values. The Transformer projects inputs to 64 dimensions, uses four attention heads and a 128-unit feedforward layer, has no positional encoding in the supplied implementation, and classifies from the final encoder position. It is evaluated on complete past windows without a causal attention mask. Table 10 supports compatibility with these two implemented differentiable score functions; it does not validate arbitrary Transformer or detector architectures.

Table 10: *Three-seed GRU and Transformer validation under the same synthetic split and controller settings*

| Detector | AUROC | AP | Full-stream refresh % | Fresh residual |
|---|---|---|---|---|
| GRU | 0.950 ± 0.005 | 0.913 ± 0.006 | 8.09 ± 0.02 | 0.00021 ± 0.00003 |
| Transformer | 0.959 ± 0.001 | 0.922 ± 0.002 | 7.44 ± 0.17 | 0.00009 ± 0.00001 |

The full-stream refresh rate is 8.09% ± 0.02% for the GRU and 7.44% ± 0.17% for the Transformer. Held-out AUROC and AUPRC are comparable in this experiment. These results support compatibility with the two tested differentiable score models; broader LSTM [22], convolutional, and production-detector coverage remains outside this evaluation.

## 7. CONCLUSION

AdaptiveDeltaXAI schedules fresh Full-SWING prediction-change maps and labels reused displays as cached event context. Three-seed full-stream temperature and power refresh rates are 8.09% ± 0.02% and 4.63% ± 0.25%. A repeated T4 batched probe shows a 6.42-fold reduction in measured explanation-stage cost under adaptive scheduling, with detector scoring and group presentation outside the timer. Public matched-count controls show higher event-refresh coverage than the tested periodic and random schedules on the documented SMD/PSM subset. Fresh-map numerical completeness and refresh coverage are distinct from attribution faithfulness: the direct 50-target perturbation comparison does not support a Full-SWING superiority claim. The evidence supports workload-specific scheduling savings, with explicit limits on cache meaning, detector quality, and generalization.

## 7.1 Limitations

Synthetic detection uses 3,436 split-local test windows, while continuous-controller coverage uses 3,456 held-out target indices. These populations contain 12 and 13 positive intervals, respectively. Full-stream operational passes contain 17,235 decisions; ablation/scaling passes contain 3,437 and retain one invalid final target only for operational summaries. Three training seeds vary model initialization; they do not provide independent synthetic data realizations.

The public evaluation covers three named SMD [2] units and PSM, not all 28 SMD machines. It includes threshold-free full-test detection metrics, exact matched-budget scheduling controls, and real fresh Full-SWING maps [1] for 50 selected windows per unit. The public study is single-seed and its calibrated threshold metrics are supplementary; it does not establish broad public-dataset faithfulness superiority.

The architecture study covers a GRU and the supplied Transformer encoder, which has no explicit positional encoding. It does not establish architecture independence, robustness across detector quality, or coverage of LSTM [22], convolutional, or production detectors.

The current-score proxy uses prefix-padded history and held-out input backgrounds; the direct test uses actual history and training backgrounds. LIME and KernelSHAP do not have equal perturbation budgets. The ablation proxy uses the first 50 fresh maps selected by each variant, and No-A5 is inactive. These limits prevent causal claims about component-level attribution quality. Exploratory direct-test intervals resample correlated windows and are not corrected for multiple comparisons.

The scaling study uses correlated feature replicas. Cached display and operational grouping have not been evaluated with operators. The controller can miss attribution changes when score and aggregate-input statistics remain small; maximum age limits persistence but does not guarantee immediate refresh. Attribution explains model behavior, not physical causes. Routine detector scoring and A5 grouping are excluded from controller timings. The supplied notebook contains execution hotfixes, and a single initialization-time code hash does not establish stage-by-stage source identity. These results are an artifact reconciliation, not a new execution of training or GPU timing.

## 7.2 Future Work

AdaptiveDeltaXAI evaluates adaptive scheduling of fresh prediction-change explanations [1] under a controlled synthetic telemetry setting and a documented public subset. Three extensions merit further study:

1. Operational grouping. The current implementation uses operational sensor groups. Future work could incorporate rack, cooling-zone, and power-distribution adjacency graphs to test whether physically grounded grouping improves operator-level interpretability and attribution utility.
2. Causal root-cause analysis. Prediction-change attribution identifies inputs associated with a model-score change; it does not establish causality. A complementary direction is to evaluate causal-discovery or Granger-style methods [23] on injected-fault streams with known source sensors, explicitly distinguishing predictive attribution from causal root-cause identification.
3. Seasonality-aware triggering. The current controller uses EMA-based score-residual processing. Future work could compare this approach with online seasonal-decomposition methods, evaluating both suppression of benign periodic refreshes and sensitivity to gradual faults.

## ACKNOWLEDGEMENTS

The authors thank the NVIDIA Academic Grant program for providing cloud computing resources and computational support for this research.

## ETHICS STATEMENT

This work uses a controlled synthetic datacenter-telemetry stream together with public SMD and PSM benchmark datasets. No proprietary production datacenter telemetry, customer data, or personally identifiable information is used. The synthetic stream contains controlled periodic and fault patterns designed to be plausible for datacenter telemetry; it does not represent a specific real facility or organization.

## REPRODUCIBILITY STATEMENT

The supplied evidence comprises a v1.9 notebook with embedded source, checkpoints, configuration, per-stage summaries, controller records, and frozen result files. The four supplied parts reconstruct an archive with SHA-256 119ffebeca5d49cb863e31679a124ee8a916c3476439685ebcee23d97574c23a, matching the supplied manifest and freeze receipt. This establishes transport integrity for this copy. The manifest records data hashes and code-tree hash 7d57cd0418187fe008ea3e9498abc9ef22848ae7752116db44374f04fa55c21b at initialization; notebook hotfixes and skip-existing behavior prevent treating that hash as a per-stage execution certificate. The author-provided project address is https://github.com/richardtrajkumar/AdaptiveDeltaXAI. No immutable public commit or archival DOI is established by the supplied materials. Reconciliation checks and remaining source-provenance limits are specified in Appendix F.

## APPENDIX A: DETAILS OF DATASET

### A.1 Physical Model

Each temperature sensor follows a baseline value perturbed by a diurnal ambient cycle (24 hour period), a CRAC driven short period oscillation, and additive Gaussian noise. Each power sensor follows a baseline value perturbed by a batch workload cycle and additive Gaussian noise. Fault events are injected as additive deviations from this baseline model according to the three regimes described in Section 3.3 and Appendix A.2. Nominal sensor baseline values and variance parameters are chosen to be plausible for commodity-server telemetry; the IPMI specification [24] informs sensor terminology rather than validating these simulated operating distributions.

### A.2 Fault Event Log

Table 11 records the author-specified synthetic fault design: 60 isolated bursts, six overlapping bursts, and six gradual drifts. These are injected design events, not a count of contiguous positive-label intervals after overlap and split truncation. Empirical event metrics in Section 6 instead use contiguous positive labels in the stated evaluated population. Dataset generation is a separate process from the v1.9 detector and explanation runner.

Table 11: *Fault event regime summary.*

| Author-specified regime | Injected events | Typical duration, steps |
|---|---|---|
| Isolated burst | 60 | 24 |
| Overlapping burst | 6 | 20 |
| Gradual drift | 6 | 75 |

### A.3 Rationale for a Single 48 Hour Dataset

All synthetic experiments use the controlled 48-hour stream rather than an additional short idealized synthetic benchmark. Its periodic structure and three injected-fault regimes provide a controlled setting; public SMD and PSM experiments are reported separately.

## A.4 Hyperparameters

Table 12: *AdaptiveDeltaXAI hyperparameters for the synthetic temperature and power experiments.*

| Parameter | Symbol | Temperature | Power |
|---|---|---|---|
| Score threshold multiplier | $k$ | 2.5 | 3.0 |
| Shared EMA update rate | $\alpha$ | 0.05 | 0.08 |
| Standardized input-drift threshold | $k_{drift}$ | 5.0 | 5.0 |
| Cached-display discount | $\gamma$ | 0.97 | 0.95 |
| Maximum display age, steps | $T_{max}$ | 60 | 60 |
| Numerical floor | $\varepsilon$ | $10^{-6}$ | $10^{-6}$ |

Table 12 records the frozen controller values. Additional fixed settings are $W = 20$, lag $L = 5$, $\delta = 1$, 64 integration segments, 100 KernelSHAP samples, and 2,000 bootstrap resamples. LIME uses its library-default sample count. Method latency has one warm-up per selected window and five repetitions. The 2 × 2 timing study has three warm-up runs and five repetitions, each with 101 decisions despite the nominal steps_per_trial value of 100. The implemented four path operands require 26 observations; the operational controller supplies 46. These roles are distinct.

# APPENDIX B: AUTOMATED VALIDATION CHECKLIST

Table 13 distinguishes checks that can be performed on the supplied artifacts from tests only recorded by the frozen precheck. A saved PASS is not a new independent execution, proof of every manuscript claim, or evidence of a public archive. No training, attribution generation, or T4 timing was rerun for this manuscript revision.

Table 13: Reconciliation checks and limits of the supplied evidence.

| Check | Evidence status |
|---|---|
| Four part hashes and reconstructed ZIP SHA-256 | Verified against supplied manifest and receipt |
| Per-seed/per-trial aggregates and displayed numeric tables | Reconciled from saved records and stage summaries |
| Target indices, cache ages, fresh residuals, display discount | Rechecked on the six saved controller archives |
| Public policy refresh counts and coverage aggregates | Rechecked from per-unit saved policy records |
| Structural, linear-completeness and alignment tests | PASS recorded in frozen PRECHECK; not rerun here |
| Mean/P95 public completeness thresholds | Rechecked on 50 saved fresh rows per unit |
| T4 environment and initialization-time source/data hashes | Recorded metadata; no independent GPU execution |
| Public immutable commit/archive and per-stage source identity | Not established by supplied artifacts |

# APPENDIX C: FULL-SWING PATH AND TEMPORAL ALIGNMENT

With $W = 20$, $\delta = 1$, and $L = 5$, the four operands use the trailing 26 observations: $X_t = x[t - 19: t + 1]$, $X_{t-1} = x[t - 20: t]$, $B^F = x[t - 25: t - 5]$, and $B^N = x[t - 24: t - 4]$, with end-exclusive slices. The controller supplies a 46-observation history $x[t - 45: t + 1]$; its first 20 observations are unused by these operands. For $\delta = 1$, $P^+ = [X_{t-1}]$ and $P^-$ is empty. The four terms are those defined before Equation (9). $J_t$ embeds a current $W \times D$ map at union rows $1, \dots, W$, while $J_{t-1}$ embeds a preceding map at rows $0, \dots, W - 1$. Their overlap is accumulated before the averaged difference is formed. The routine returns a history-shaped array; the compact map $E_t$ denotes its trailing $W + 1$ rows. All earlier rows have zero attribution.

Each piecewise-linear path uses 65 uniformly spaced nodes over its entire path parameter interval and sums trapezoidal gradient averages times consecutive node displacements. The current-far path has two linear legs for $\delta = 1$; the other three paths have one. The node count is per complete path, not 64 segments per leg. Batched evaluation stacks nodes within each term; the serialized timing arm evaluates identical nodes individually. Completeness is approximate for nonlinear models and is checked on raw fresh maps.

# APPENDIX D: RESULT WORKLOAD MAP

The population and state policy differ by result family. Table 14 records those distinctions. Direct difference attribution uses actual pre-target history, including pre-test observations when needed, with training-only normalization. The current-score proxy instead constructs padded history. Neither the full-stream controller nor its carried state is used to report held-out detector quality.

Table 14: Workload populations and controller-state policies.

| Result family | Population / selection | Seed and state scope |
|---|---|---|
| Synthetic detection | 3,436 windows; 331 positives; 12 intervals | 42, 52, 62; split-local windows |
| Ablation and scaling | 3,437 operational decisions; final target invalid; proxy on first 50 fresh maps | 42, 52, 62; state initialized at endpoint 13,843 |
| Refresh coverage | 3,456 held-out targets; 13 positive intervals | 42, 52, 62; state carried from full stream |
| Full-stream controller / architecture | 17,235 decisions; final target invalid | 42, 52, 62; endpoint 45 onward |
| Current-score proxy | 50 common positive test windows; padded history | 42; sampled held-out backgrounds |
| Direct difference perturbation | 50 common positive test targets; real history; 21 × 8 union | 42; 30 training-union backgrounds |
| Fresh-method timing | 10 windows × 5 repetitions = 50 trials / method | 42; one warm-up/window; padded history |
| Batching × scheduling | 101 decisions / repetition; adaptive 17 refreshes | 5 repetitions; 3 warm-up runs; state reset |
| Public scheduling / fresh subset | Full public test scores / 50 selected positive maps per unit | 42; retrospective exact-count controls |

# APPENDIX E: PER-UNIT PUBLIC POLICY OUTCOMES

Table 15 gives per-unit public coverage underlying Table 9. Every policy has the same realized adaptive refresh count within a unit, selected retrospectively after seeing the completed adaptive schedule. Periodic and random schedules do not use future scores after count and horizon are fixed, but their budgets are not prospective online commitments. Random results use one seed-42 schedule; the score-change oracle uses the full score sequence.

Table 15: Per-unit public event-refresh coverage at exact matched counts.

| Dataset / unit | Events | Count | Adaptive | Periodic | Random | Offline oracle |
|---|---|---|---|---|---|---|
| SMD machine-1-1 | 8 | 1,643 | 8/8 | 6/8 | 6/8 | 8/8 |
| SMD machine-1-2 | 10 | 1,399 | 10/10 | 9/10 | 10/10 | 10/10 |
| SMD machine-1-3 | 12 | 1,388 | 12/12 | 10/12 | 10/12 | 12/12 |
| PSM | 72 | 4,832 | 54/72 | 38/72 | 29/72 | 59/72 |

# APPENDIX F: Reconciled Attribution and Evidence

## F.1 Direct Prediction Difference Evaluation

For a method-specific ranking, $M_k$ replaces the same absolute coordinates in each overlapping window with the normalized-zero reference. The primary perturbation measure is

$$AOPC_{\Delta}(t;K) = \frac{1}{K}\sum_{k=1}^{K} \quad [|\Delta s_t| - |f(M_k X_t) - f(M_k X_{t-1})|] \quad (24)$$

Table 16: Direct prediction-difference perturbation, 50 common temperature targets at seed 42. Individual and paired intervals are exploratory window-bootstrap 95% intervals (2,000 resamples). The paired column is Full-SWING minus the row method.

| Method | Mean | Window 95% CI | Paired difference [95% CI] |
|---|---|---|---|
| AW-IG | 0.0061 | [-0.0414, 0.0533] | -0.0215 [-0.0641, 0.0183] |
| Full-SWING [1] | -0.0154 | [-0.0588, 0.0309] | Reference |
| KernelSHAP [5] | 0.0261 | [-0.0103, 0.0677] | -0.0415 [-0.0820, -0.0078] |
| LIME [4] | 0.0106 | [-0.0154, 0.0453] | -0.0260 [-0.0576, -0.0009] |
| Zero-baseline IG [6] | 0.0236 | [-0.0149, 0.0697] | -0.0390 [-0.0768, -0.0064] |

## F.2 Quadrature and Grouping

Let $u_0, \dots, u_N$ be nodes on a complete piecewise-linear path, with $N = 64$. The supplied implementation computes its native $W \times D$ map as

$$IG_N = \sum_{j=0}^{N-1} \frac{\nabla f(u_j)+\nabla f(u_{j+1})}{2} \odot \left(u_{j+1} - u_j\right) \quad (25)$$

The componentwise product integrates along the sampled path. Equation (9) embeds each term on absolute coordinates before subtraction. For a display $C_t$ and a fixed channel group $g$, the separate presentation output is

$$G_{t,g} = \sum_{d \in g} \sum_{l \in H_{t_{last}}} C_{t,l,d} \quad (26)$$

Grouping sums signed contributions and may cancel opposing channel effects. The supplied temperature groups divide the channel list in half; neither group is a measured thermal zone. No-A5 does not alter the executed controller, so the frozen ablation provides no empirical grouping-effect estimate. Here the temporal union is that of $t_{last}$, the map's originating refresh endpoint, including on later cache-hit displays.

## F.3 Public Fresh Map Completeness

Table 17: Public fresh-only completeness residuals from 50 selected positive windows per unit, seed 42. Gates are $mean \leq 0.05$ and $P95 \leq 0.15$ in the detector's log1p-MSE score units. Passing these gates does not establish perturbation faithfulness.

| Dataset / unit | Fresh maps | Mean residual | P95 residual | Gate |
|---|---|---|---|---|
| SMD machine-1-1 | 50 | 0.002999 | 0.008568 | Pass |
| SMD machine-1-2 | 50 | 0.001499 | 0.000975 | Pass |
| SMD machine-1-3 | 50 | 0.000241 | 0.001044 | Pass |
| PSM | 50 | 0.000016 | 0.000087 | Pass |

## F.4 Reconciliation Scope and Source Provenance

This revision checks each numeric table against its declared saved result family, recomputes seed/trial aggregates and method-latency percentiles, and cross-checks the six controller archives for target alignment, cache ages, display discounting, refresh counts, and raw-fresh completeness. It checks the direct perturbation means and paired window-bootstrap intervals against per-window rows, and public counts/coverage against per-unit policy records. Empirical figures are reproduced from controller arrays, frozen curves, and summary rows. No detector training or fresh GPU attribution is rerun; saved checkpoints without the original raw arrays do not independently re-establish detector AUROC/AP.

The notebook embeds its source archive and supplies patches removing a missing controller import, matching cache-fidelity history to the 46-observation controller history, and correcting the controller-record loader. Another cell concatenates the five method timing files before post-hoc analysis. These patches explain the supplied execution path, but the initialized code-tree hash and skip-existing stage behavior do not certify the exact source of every already-present stage output. An immutable source revision with the patches integrated, plus a public archive identifier, is still required to claim a fully pinned public release. This is a source-provenance limitation, separate from the verified checksum of the supplied results.

## APPENDIX G: Supplementary Public Calibration

The supplementary detector fits and normalizes on the first 80% of the public training partition, then sets a threshold at the 99.5th percentile of scores on the remaining chronological 20%. It is a separately trained checkpoint, so its AUROC/AP need not equal Table 8. Test labels are used only for evaluation. Table 18 retains the uneven thresholded outcomes rather than implying that threshold-free ranking quality guarantees useful alerts.

Table 18: Supplementary public calibration, seed 42. These checkpoints and thresholds differ from the primary threshold-free public detector protocol.

| Dataset / unit | AUROC / AP | Precision | Recall | F1 | Events detected |
|---|---|---|---|---|---|
| PSM | 0.760 / 0.550 | 0.939 | 0.004 | 0.008 | 24/72 |
| SMD machine-1-1 | 0.876 / 0.527 | 0.192 | 0.982 | 0.321 | 8/8 |
| SMD machine-1-2 | 0.874 / 0.213 | 0.324 | 0.066 | 0.110 | 9/10 |
| SMD machine-1-3 | 0.587 / 0.146 | 0.972 | 0.043 | 0.082 | 1/12 |

PSM point recall is 0.0038 and only 24/72 events are detected under this threshold; SMD machine-1-3 detects 1/12 events. These outcomes limit operational alerting claims. Explanation refresh coverage in Tables 9 and 15 describes score-only scheduling on the primary public score sequence and must not be substituted for calibrated detector event recall.